\documentclass[english,aps,pra,reprint,noshowpacs,superscriptaddress,]{revtex4-1}  
\usepackage[T1]{fontenc}	
\usepackage[latin9]{inputenc}	
\usepackage{geometry}		
\usepackage{graphicx}
\usepackage[above,below]{placeins}	
\usepackage{times}
\usepackage{hyperref}  
\hypersetup{colorlinks=true,urlcolor=blue,citecolor=blue,linkcolor=blue}   
\begin{document}

\title{A two-phase study examining perspectives and use of quantitative methods in PER}
\author{Alexis V. Knaub}
\affiliation{Department of Physics and Astronomy, Michigan State University, East Lansing, MI, USA 48823}
\author{John M. Aiken}
\affiliation{Center for Computing in Science Education \& Department of Physics, University of Oslo, N-0316 Oslo, Norway}
\author{Lin Ding}
\affiliation{Department of Teaching and Learning, The Ohio State University, Columbus, OH, USA 43210}

\begin{abstract}
\noindent
While other fields such as statistics and education have examined various issues with quantitative work, few studies in physics education research (PER) have done so. We conducted a two-phase study to identify and to understand the extent of these issues in quantitative PER . During Phase 1, we conducted a focus group of three experts in this area, followed by six interviews. Subsequent interviews refined our plan. Both the focus group and interviews revealed issues regarding the lack of details in sample descriptions, lack of institutional/course contextual information, lack of reporting on limitation, and overgeneralization or overstatement of conclusions. During Phase 2, we examined 72 manuscripts that used four conceptual or attitudinal assessments (Force Concept Inventory, Conceptual Survey of Electricity and Magnetism, Brief Electricity and Magnetism Assessment, and Colorado Learning Attitudes about Science Survey). Manuscripts were coded on whether they featured various sample descriptions, institutional/course context information, limitations, and whether they overgeneralized conclusions. We also analyzed the data to see if reporting has changed from the earlier periods to more recent times. We found that not much has changed regarding sample descriptions and institutional/course context information, but reporting and overgeneralizing conclusions has improved over time. We offer some questions for researchers, reviewers, and readers in PER to consider when conducting or using quantitative work.
\end{abstract}
\maketitle

\section{Introduction}
Quantitative research can provide valuable insights and be compelling for audiences as they often draw upon large sample sizes and use compelling statistics \cite{BeichnerGS}. In PER, quantitative work is typically employed to provide numeric ``... observations through some statistical techniques in order to better describe, explain, and make inferences about certain events, ideas or actions in physics education'' \cite{DingGS}. However, the usefulness of this work is only as good as its quality. Quantitative work is not immune to issues regarding data collection and analysis, as well as researcher bias \cite{Typology}. The research in others fields indicates that peer-reviewed quantitative research has room for improvement in how current techniques are used. While few of these studies were within physics education research (PER), it is likely that PER also has areas ripe for improvement. However, without investigation, we cannot assume that PER has the same set of quantitative research issues that other fields have. 

To better understand the issues within peer-reviewed quantitative PER,  we sought to find out what these issues are. This study has the following research goals:
\begin{enumerate}
\item Identify issues in quantitative physics education research as well as ways to improve via advice from experts
\item Determine how pervasive identified issues are in PER
\end{enumerate}

Our first goal was used to identify which issues we should focus on in order to create a cohesive project that has relevant research questions and results. Our second goal was to understand the extent of identified issues. 

We addressed these goals in a two-phase study design. In the first phase, we asked quantitative PER experts for their perspectives on the biggest issues that quantitative PER has. In the second phase, we examined peer-reviewed articles on assessments from the American Journal of Physics (AJP), Physical Review- PER (PRPER), and The Physics Teachers (TPT) to determine the extent to which these expert-identified issues exist. 
 
\section{Background and motivation} 
\subsection{Research design issues}
Regardless of whether a study is quantitative, a research project's design determines the course of a research project, including which data are collected, how data are collected, and what analyses can be done. Research design decisions are often determined by the research questions that a researcher seeks to answer \cite{DingGS}. Some decisions include which data to collect and whether the data are considered categorical, continuous, etc. \cite{Fields}. What is measured should have some understood theoretical backing. Scott (2011) emphasized ``Only if a researcher has a clear understanding of the logic of a particular measure can he or she make an informed sociological judgment about its relevance for a particular piece of research'' \cite{Scott}. If care is not taken, the results may lack validity (i.e., whether an instrument measures what it claims) and reliability (i.e., whether the instrument's questions are consistently interpreted) \cite{Fields}. More information on nuances of reliability and validity (e.g., the importance of multiple sources of evidence for claims) found in many standard textbooks (e.g., \cite{Creswell}).

Besides considering the research questions and how the design answers those questions, it is also important to consider limitations within the study's design and what is truly being asked. Docktor \& Mestre pointed out that PER studies that use cognitive psychology often occur within a lab and thus, the results may be different if one tries to replicate the study in a classroom \cite{Docktor}. Another example is who is included in the sample. While some decisions regarding who is included in the sample are deliberate (e.g., purposive sampling), others may be accidental because a portion of the population would not be identified (e.g., only relying on who is listed in the phonebook) even if they are a part of the population of interest (e.g., convenience sampling) \cite{Fowler}. Some research questions may be answerable, but their answers are not meaningful (e.g., comparing two groups for no apparent reason) \cite{VA}. 

\subsection{Analysis issues in quantitative work}
Quantitative work can have issues beyond how the research project is conceptualized. Literature indicates that studies in many fields, such as education and medicine, may violate assumptions built into these statistical models or researchers may incorrectly interpret the statistics (e.g., making causal claims when one cannot). Examples include incorrect use and interpretation of: value added modeling \cite{VA},  multivariate statistics \cite{Multivariate}, exploratory factor analysis  \cite{EFA}, and p-values \cite{Pvalue, SigTest}. Parametric statistics, such as t-tests, assume that the data have a normal distribution and are interval data \cite{Fields}. The examples given are not just hypothetical but have been observed by researchers in various fields. There have been some studies regarding statistical issues in PER, such as issues with normalized gain (e.g., \cite{NGain1, NGain2}).

If statistics are not carefully used and reported, some issues can arise. Misinterpretation of p-values may mean that result seems much more important than it is \cite{Pvalue, SigTest, EffectSize}. To better interpret results, effect size and confidence intervals should be reported \cite{Pvalue, SigTest, EffectSize}. Some of these issues have led to false positives and false negatives (i.e., type 1 and type 2  errors) \cite{FalseNeg}. In other words, these errors can mislead researchers to believing their results have statistical significance when they do not (i.e., false positive) or that the results lack statistical significance when they do (i.e., false negative). 

Other issues are more general. Researchers may not have information on all individuals in a sample, resulting in missing data. For example, surveys have response rates that indicate what percentage of individuals in a sample took a survey. If only a particular group of individuals took a survey, biased results that do not accurately portray the population are possible \cite{Fowler, MissingData}. Researchers may also make decisions on which cases to keep in their data. ``Cleaning'' data or selecting cases that do not seem valid can introduce errors, especially if removed cases were actually valid data \cite{Typology}. How data are aggregated matters, because important information can be lost. For example, the percentage of postsecondary degree holders has large variance among different Asian American ethnicities (e.g., Chinese, Vietnamese) \cite{Disaggregate}. When data are reported in aggregate, these details are lost and mislead audiences to believe Asian Americans are universally succeeding \cite{Disaggregate}.

Graphics can be useful, but they also can portray data inaccurately \cite{Fields, Typology}. One such example is a histogram may indicate a normal distribution, though calculations indicate the distribution is not normal \cite{Typology}. Another such example is that a graph's scale may distort the data to de-emphasize differences between two groups \cite{Fields}.

Clearly defining terms is important as words can mean different things \cite{Scott}. In a study on equity, how  ``equity'' is defined and what the study intends to examine can lead to different interpretations of data \cite{EquityModel}. The authors advocate that researchers are explicit in their definition of equity. While this paper focused on one term, such explicitness is likely useful for other concepts and generally interpreting results. Ding and Liu (2012) noted the following:
\begin{quote}
However powerful a statistical analysis may be, it is after all just a tool that can inform one of what a result is but cannot tell why the result is such. It is the researcher's job to make credible inferences for the reasons that underlie the results and connect them back to the original theoretical framework'' \cite{DingGS}.
\end{quote}

Quantitative research has its limitations in what can be described and thus, claims should be carefully made. While statistics can describe trends within large groups, understanding what an individual is thinking may not be possible \cite{BeichnerGS}. Correlation does not equal causation, as many factors can produce high correlations \cite{Typology, DingGS}. Data depict some student attributes and are taken during a specific period \cite{LAEthics, LAEthics2}. If these data are to be used to guide decisions for students, such as recommending which courses they should take, researchers suggest that those using the data keep in mind that students are individuals and that a student's past does not necessarily determine their future  \cite{LAEthics, LAEthics2}.

\subsection{Motivation}
In summary, the broad literature on research design and analysis indicates there are multiple potential issues, ranging from how data are collected  to whether statistics are used appropriately to how they are then interpreted. Some of these studies are in PER. Yet, it is unclear how pervasive these issues are or whether these issues exist. Our goals with this paper are to examine which issues exist in PER and to what extent they do. 

\section{Phase 1: community perceptions on quantitative physics education research}
\subsection{Research questions}
The research questions for Phase 1 are as follows:
\begin{enumerate}
\item What issues exist in quantitative PER, and how are the issues ranked by experts? 
\item What suggestions do experts have to improve the robustness of quantitative PER? 
\end{enumerate}
Experts for this study were identified by the editors of AJP and PRPER. 

\subsection{Methodology}
\subsubsection{Focus group sample description and protocol}
We asked the editors of AJP and PRPER for some suggestions of whom they consider experts in quantitative PER and would be interested in participating in focus group regarding quantitative PER; we specified that these individuals do not necessarily need to work exclusively in PER or exclusively with members of PER but have enough familiarity to be able to comment on quantitative PER. 

The total list included 27 individuals. While we believe that the editors and advisory board did indeed identify quantitative experts in PER, we also anticipate that this list is not comprehensive; we do not want readers to believe there are only 27 quantitative experts in PER. For the focus group, we contacted 8 individuals.  We wanted a diverse set of opinions and did not want the focus group to reflect a particular research tradition, group, or advisor. Individuals were selected based on current institution affiliation; where they received their doctorates and did postdoc work (when applicable); and who recommended them. Institution affiliations and degree information were gathered from group, departmental, or personal websites. 

Five individuals agreed to a day and time to meet virtually through a video meeting. The other three declined due to time commitments. Three people attended the focus group conducted by Aiken and Knaub. The three who attended all were from different institutions and had different doctoral and postdoctoral backgrounds. They represent the recommendations of two editors or journal advisory board members. 

As this was a focus group, we used a semi-structured protocol aimed at generating discussion among the focus group participants. We included these questions in Appendix A. Questions delved into participants' general opinions on quantitative PER, challenges and mistakes within quantitative PER, how pervasive these issues are, and recommendations for resources. Based on the literature and our experiences, we believe there are issues within quantitative PER. However, we did not want to lead the focus group to confirming our beliefs hence asking open questions.
 
\subsubsection{Interview sample description and protocol}
Based on the  focus group's feedback, we developed a project plan for Phase 2. To refine the project plan, we conducted interviews for feedback and other suggestions. We contacted seven individuals, purposively selecting for diversity in current institution, doctoral institution, postdoctoral institution (when applicable), and recommender. 

Six agreed to be interviewed, including two individuals who were supposed to attend to the focus group. Only one individual declined, claiming not to be an expert in quantitative PER. The individuals we interviewed represent the recommendations of four  editors or journal advisory board members and a variety of backgrounds.

\subsubsection{Limitations and threats to validity}
For this phase of the study, our limitations and threats to validity are the small size of the focus group and presenting a pre-made plan. For the former, perhaps a larger or different set of individuals would have identified other issues in quantitative PER. However, our follow-up interviews indicated these issues are present in quantitative PER.

Regarding a pre-made plan, perhaps, had our interviewees not seen this plan, other issues may have been identified.  To make sure we received candid feedback, both interviewers took care to ensure that interviewees felt comfortable critiquing the plan and making other suggestions if they thought there were other issues we should focus on. We used broad, open-ended questions and encouraged interviewees to be honest and offer alternatives if they felt we were focusing on a non-issue. We also anticipate that there may be other issues in quantitative PER and hope that other issues in research are explored in the future.

\subsection{Results: identified issues in PER from the focus group}
We present findings from our focus group.  Individual focus group participants are referred to as F(number) (e.g., F1).

Our focus group identified assessments, such as concept inventories and attitude surveys, as one of the primary areas where PER has focused its quantitative work. Literature has indicated that assessment, particularly multiple choice concept inventories (e.g., the Force Concept Inventory) historically has been and continues to be an important part of PER \cite{Docktor}. In terms of specific issues, the focus group identified several issues within quantitative PER. While they did discuss issues such as p-values and effect sizes, they primarily focused on the following:
\begin{itemize}
\item \textbf{Not reporting on limitations well.}  If care is not taken to report on a study's limitations, our focus group thought readers may come to incorrect conclusions. One focus group participant, F1, explained this: 

\begin{quote}If you're not careful about your presentation, the unwary reader can take messages away from quantitative papers that are not valid because of the type of data that you have, the type of techniques that you used, or that kind of thing. So, I think reporting on the limitations of statistical techniques is something that we don't do well enough, sort of putting boundaries on the realm of applicability of our findings, both from a demographic point of view, as F2 mentioned, but also from a statistical point of view.

\end{quote}
 F3 gave an example of a limitation within quantitative work: 
 \begin{quote}I think a lot of questions, if you crudely characterize questions that are about how or about mechanisms, are a lot harder to answer quantitatively. Like, how do students approach this new set of inquiry materials, you're never going to get a good answer to that quantitatively unless you find some extremely innovative ways of quantifying the student experience, and I have trouble with imagining how that would work.
 \end{quote}
 F3 explained that in this example, qualitative work could complement the quantitative. However, F3 was clear that the quantitative by itself would likely have limitations on what was found.

\item \textbf{Institutional/course contexts and samples are not described well.} This was succinctly stated by F2: ``PER does not do a good job of telling its audience who it is that they're studying.'' Describing one's population and sample is important because errors can result if a study's implications are applied to a completely different population or if the sample is not representative of the population \cite{Fowler}. F1 pointed out that these descriptions need to be done thoughtfully or the audience might get lost: 
\begin{quote}So, it's a balance, I think when you're publishing, to try and provide enough information that people can really understand what you're doing, but not more information... [that] ultimately make the paper unreadable or so they can't understand it if they're not already invested in the literature and the statistical techniques being used.
\end{quote}
 
\item \textbf{Overgeneralizing and overstating results.} F2 remarked that ``the assumption is that if I present data about my students, it's going to be equally valid for your students.'' The idea is that context varies in critical ways (e.g., student backgrounds, resources available) that may impact whether one can expect similar results in a completely different context. F1 concurred, pointing out that if authors are not careful, readers may draw inaccurate conclusions: 
\begin{quote}I think we tend to make conclusions based on these findings that are sometimes not entirely valid. So, it's not necessarily I think that the statistical techniques themselves are flawed, but if you're not careful about your presentation, the unwary reader can take messages away from quantitative papers that are not valid because of the type of data that you have, the type of techniques that you used, or that kind of thing.
\end{quote}
\end{itemize}

There were a few areas where the focus group identified potential issues but also saw value in current practices, even if they are imperfect. Some examples include:

\begin{itemize}
\item \textbf{Implicit theories}. Some papers do not describe the theory (e.g., a theoretical framework that uses social, cognitive, and/or learning theories) that guides the study. The focus group believed that rather than lacking any theory, there are implicit theories that authors may not articulate. F1 explained why this could be an issue: 
\begin{quote}If somebody comes into that data with a different perspective than was intended by the authors or a different idea of what learning looks like than what was intended by the authors, they can very easily misinterpret what those data are saying.
\end{quote}

However, when the idea of requiring theories to be explained came up during the discussion, the focus group was divided. While F1 saw ``...no harm can come from articulating it, and being forced to articulate it'',  F2 was hesitant to make such a stance, stating:
\begin{quote}I worry about making that a requirement of publication, and this has been on my mind because there are people that are really pushing to have that be a requirement of publication... If I look at something from the University of Washington, I sort of know what their framework is because the papers that they've published before all sort of come from the same place. It doesn't strike me as necessary to reiterate that at the beginning of every paper.
\end{quote}
F2 was not opposed that peer reviewers suggest explicitly articulate theories but against authors being forced to articulate theories. F3 was in the middle: 
\begin{quote}I wish there would be a nice middle way because I feel like for those researchers, if they don't use some of their writing time to help think through the theoretical basis for their work, then they're also missing out on a chance to grow.
\end{quote}

\item \textbf{Not thoroughly considering all aspects of research design and articulating the research design.} Our focus group pointed out the level of detail other fields spend on the research design in terms of data collection, hypotheses, etc. F2 explained this: 
\begin{quote}F1 talked about earlier, but the care that's gone into the design of the experiments, the planning of how many students we will need to get a statistically significant result. Very carefully stating what the goal of the experimenter is, what a null result would look like, etc.
\end{quote} 
However, the focus group was cautious to suggest that PER adopts these practices as a universal standard for all manuscripts. The potential dangers of a more stringent standard were articulated by F2: 
\begin{quote}One thing I worry about is a lot of times, people aren't looking at things as this can be a big tent. It's like they didn't do this and so their research is not valid or useful, and I think that's a danger for PER. Just because it's not the way you want to do research, doesn't mean that it's not useful ... That [the research] isn't necessarily compelling or isn't teaching us something.
\end{quote} 
Depending on what standards were adopted, there is potential some research simply could not meet the standards. F3 pointed out that such studies may not have made mistakes but have constraints that pose a limitation: 
\begin{quote}If you're a researcher at a smaller institution, you can't get some variation between the classes that you're studying, so you're limited by what you can do but you still want to contribute to the enterprise of PER so you're going to do what you can do.
\end{quote} 
\end{itemize}

Despite the acknowledged issues, the focus group believed quantitative work in PER is improving, becoming more nuanced and following the quantitative practices adopted by other fields that are believed to be good practices. They believed that because PER is a newer field, researchers did not initially use as many statistical techniques. F1 pointed out some statistical techniques are newer to PER but are likely to become part of quantitative practice: 
\begin{quote}I think as we start to ask more sophisticated questions, reporting of effect sizes, the importance of effect sizes, as far as the value of the research, and then conversations about statistical power from the beginning, I think, are going to become more important. They're not something that I've seen a lot in the literature as of yet, but I think they're going to be important in the future.
\end{quote}

At the same time, while PER may be moving in this direction, the focus group, as seen in the previous findings, saw room for a variety of research papers. F2 emphasized a ``big tent'' where there is room for these more detailed papers as well as papers that do not go into these details. They suggested that other authors can publish manuscripts that critique work that lacks certain elements (e.g., an explicit theoretical framework).

\subsection{Results: feedback on project plan from interviewees}
Overall, the interviewees thought our project plan for Phase 2 would be useful for PER. The project plan reflected the interviewees'  concerns and observations. Some of their comments expanded upon the themes of the focus group, pointing out more detailed issues and ramifications if quantitative work is not done well. They also pointed other issues within PER that they observed. Below we summarize the additional insights that the interviews provided, linking some of the interviewees' comments to the overarching themes from the focus group as well as summarizing other issues they mentioned.  Individual interviewees are referred to as I(number) (e.g., I1).

\begin {itemize}
\item \textbf{Not reporting limitations well}
\begin {itemize}
 \item \textit{Variability with individuals in studies.} Two interviewees described two limitations regarding studying people. One is that while samples can be representative of a given population, there is variability. I2 pointed that ``we're not certain that the results are generalizable... This is really a complex system that we're looking at, or complex systems.'' I3 gave a specific example: 
 \begin{quote}Sometimes you see studies... where there are direct comparisons drawn from treatment and control groups where the treatment group is at one high school and a control group is at a different high school. You just can't do that. This is highly, highly problematic. On top of that, when you do those kinds of things, once again you fall back to the presumption that there is an inherent similarity across individuals that just fundamentally doesn't exist. It just isn't there.\end{quote}
 
 Among individuals, even with similar demographic markers (e.g., same race, gender), there is variability. I3 emphasized that:
 \begin{quote}People are not electrons, nor are they protons, nor are they photons. There isn't a degree of similarity between them. There is a variety of difference from individual to individual, and as a result, making these kinds of broad-based claims based on statistical analyses is something that should be done with care, if done at all.
 \end{quote}
 
 I3 also suggested another issue with variability with individuals, that individuals themselves are not necessarily consistent from one day to the next by chance:
 \begin{quote}
 When you look at FCI, when you look at these other instruments that you have listed here, CSEM, BEMA, CLASS, it's important to bear in mind that if you give the instrument to the same individual the next day, the responses might not be identical. That is something to accept, think about, and contend with as a researcher. You can't say, well, this is what it is. There is some degree of error associated with that.
 \end{quote}

\end{itemize}
\item \textbf{Institutional/course contexts and samples are not described well.}
\begin{itemize}
\item \textit{Missing data} Two interviewees discussed how PER may not handle missing data well. I4 simply stated ``There's almost never  any discussion of [missing data]. Not even the information to be able to know if there is missing data.'' I1 believed that when researchers account for missing data, they tend to do paired analysis (e.g., only include students who took both the pre- and post-tests). While I1 felt this could be a step in the right direction, they also thought that paired analysis has some critical limitations: `` But you haven't accounted for the people that were, that sort of, were invisible, right?'' I1 advocated for using imputation methods to examine missing data:
\begin{quote}If you did an imputation method you would see the pattern. You would see, "Oh, students who, over on this question, identified as female also systematically don't answer this question. Hmm maybe there's a problem there." And then you can basically fix it through some imputation method. Or at least you can account for the increased ignorance from not having those responses. \end{quote}
In sum, analyzing missing data can offer additional insights that may not be apparent otherwise.
\end{itemize}
\item \textbf{Overgeneralizing  and overstating results.}
\begin{itemize}
\item \textit{Making causal claims when one cannot.} I3 was concerned with how researchers in quantitative work make causal claims: 
\begin{quote}My primary concern has to do with this idea that somehow performing an empirical study or an experimental study of some kind where there is a direct comparison between a treatment and control group allows you to make a causal claim. Fundamentally, in educational research of all types that would be a huge mistake to do that. It has a lot to do with the idea that individuals from person to person vary a great deal, and that's generally accepted... I see these kinds of claims, and this is a much more consistent kind of problem in a lot of papers that I've reviewed over the years... Even if you have an experiment, to make a direct comparison, to find a statistical difference between two groups and then on top of that to say, `Well, then as a result, this shows [the treatment worked],' that's very dangerous territory to go into.
\end{quote}

 I3 emphasized they were not trying to stop people from doing education research or making claims but that I3 was encouraging good research practices:
 \begin{quote}There's a difference between not being able to do it at all and not being able to draw these overarching causal conclusions and being careful about what we say and being thoughtful about how we make certain kinds of claims in what it is that we found. Making definitive claims in educational research is fraught.
 \end{quote}
 
 \item \textit{Implicit ungeneralizability.} I1 hypothesized: 
 \begin{quote}[Authors may] really only care about their students, for example, at their institution. And so they may be very implicit about the fact that they really aren't speaking to anything beyond the edges of their campus. So you sort of have to, I think, read that carefully, in some sense. That's also a slippery slope because sometimes people forget that their work is very institutionalized and it doesn't apply to other schools.
 \end{quote} 
 According to this interviewee, explicitness regarding context and who was studied would help audience understand that the work may not apply to all institutions. 
\end{itemize}
\item \textbf{Implicit theories.}
\begin{itemize}
\item \textit{Lack of theoretical basis for interpreting measures.} I1 stressed the importance of theory with quantitative PER: ``Does the measure make any sense, right?'' Without having any theory to support the study design or claims, measurements may lack meaning. 

\end{itemize}
\item \textbf{Not thoroughly considering all aspects of research design and articulating the research design. }
\begin{itemize}
\item \textit{Statistics not being used thoughtfully.}  I4, when reflecting on PER's past and current use of statistics, said ``There are more statistics, but I don't know if that's made it a lot better off than when there were very few statistics.'' I3 warned that particular statistical methods that become popular may be treated as the only method: 
\begin{quote}There's this idea that certain kinds of statistical methods work better than others and that kind of thing. For example, right now in quantitative research you see HLM [hierarchal linear modeling] being thrown around as if it were some kind of magic amulet of you hold it up, ooh, HLM. It's not the statistical method, it's the analysis that's important. You do your data collection, you do the work, and sometimes hierarchical linear modeling works and it's what you're supposed to do, and other times it's better just to go with an ANOVA or a multiple regression.
\end{quote} 
I5 reiterated this point, emphasizing that ``[statistics] are a tool and given your specific use, not all wrenches are the same.''

One issue identified by two interviewees is that researchers in PER sometimes do consider the meaning of their results in the particular context. I5 pointed out that blanket criteria that are applied to all studies do not work: 
\begin{quote}I feel lots of people are using methods from the shelf, and they're believing in just some criteria. So it's almost like people believe that if reliability is not point seven, it's a shitty instrument, and then once it's point seven, it's a super instrument. And that's just not the case... It depends, right?
\end{quote}
I6 pointed that when in interpreting statistics, a result can be statistically significant but not necessarily meaningful: 
\begin{quote}So, is a difference of half a point on an assessment for 10,000 students, is that really significant or not? The statistics might tell you it is, and you look at it and you say, "naaahhhh, no. It really is not." I think sometimes people just throw the statistics out there and don't think enough about what it's really telling them, or not telling them.
\end{quote}
\end{itemize}
\end{itemize}

During the interviews, the idea of minimum criteria or standards for reporting data in quantitative manuscripts came up. The interviewees were not entirely opposed to minimum standards. I5 stated that ``what we are lacking is standards in what we would expect from a paper on the reliability and validity of instruments, on certain procedures, it seems like there's different standards.'' I3 believed this was important for PER as a field: 
\begin{quote}
Emerson once said many years ago, he says, ``Look, it is not the direction that you're going in any given moment but the general trend that you go in over time that is the most important thing to consider about one's life.'' That is in a sense what we're trying to do about research. It is not what the individual finding of a particular paper might be but the general trend of the field over time that we happen to go in. I think that if we identify what's clear in terms of good methodology for these types of empirical/experimental studies, that we will have a better chance of finding these kinds of trends that lead us to the kinds of conclusions that help us improve physics education for the long term.
\end{quote}
However, similar to the focus group, the interviewees saw this as nuanced issue with potential challenges if standards were imposed. Although not opposed to minimum standards, I2 expressed concern regarding reporting one's sample: 
\begin{quote}
I think there needs to be a balance of how much we need to say about the population. Because there could be some ethical problems going ahead. That being said, I think it's important and I personally strive to try and say, `Okay, so what is this kind of population that we have here? And how would it be different from other places?' 
\end{quote} 
One of the ethical issues noted by I2 is that too many demographic details might de-anonymize the sample. I5 also pointed out that because PER is a developing field, whatever standards might be created will shift with time as knowledge bases grow.

One aspect that a few interviewees brought was the culture (i.e., beliefs, practices, etc.) around quantitative research in PER. I2 suggested that many in PER have a narrow definition of what quantitative research is: 
\begin{quote}When you say quantitative methods the thing that a PER person thinks is, the Force Concept Inventory, or the CLASS. They don't think ... I'm using statistics and counting things. The very specific realm of what it means in PER, which is not actually what it means in other fields.
\end{quote} 
I6 observed that some individuals may not use quantitative work possibly due to a misunderstanding of what quantitative work is: ``I've seen people that will not apply quantitative methods when maybe they should because they think their population has to be super duper huge in order to do anything meaningful, which again, is I think a lack of knowledge about things beyond mean and standard deviation, and Z scores.'' I4 observed that there are few replicability studies, which they feel can be problematic if a single study is used to support a claim. I1 noted that the way articles are written in education fields differ from PER, perhaps as a result of how traditional science papers are written: 
\begin{quote}
``[Articles in journals such as Journal of Research in Science Teaching (JRST)] always, like, have the million-page introduction to every paper. And that style is pretty different than PRPER's, even today... Because traditional science papers are often written in that very staccato style, where if you read a Science or a Nature or a PRL article in a traditional science discipline, right? They'll just say, `Here's a two-paragraph summary with 50 citations that, if you want to understand this topic, here's the 50 things.' Whereas JRST will say, `Well let's talk about what this all means.' And it's partly, I think, intended to be a little more self-contained.  I come from the science background where I'm perfectly happy to give a short introduction, but I've learned that this is sort of a stylistic thing and a rhetorical flourish. But it does mean that it's hard to do atheoretical work and publish it in JRST, or Science Ed, or similar journals. 
\end{quote}
While some of these cultural differences may be actual problems (e.g., not using quantitative methods due to lack of understanding), other differences, such as the differences in introductions, are areas that may be worthwhile thinking through the costs and benefits. Considering the cultural aspects regarding quantitative PER may be useful for considering how any identified issues could be changed.

\subsubsection{Concerns regarding Phase 2}
A few interviewees brought up some potential issues with our plan for Phase 2. I1 pointed out that American Journal of Physics (AJP), The Physics Teacher (TPT), and Physical Review- PER (PRPER) all have difference audiences and thus, authors would present their work differently. The rationale will be described in more detail in Phase 2, but AJP and TPT were traditionally the venues for PER prior to PRPER. I1 also brought up a subtle point, whether we would check to see if claims are consistent with the data that is presented or whether we would check to see if they answered the research questions or goals. We opted to do the former. Answering the broader research questions is important, we are more interested in more fundamental research practices, as our focus group and interviewees suggested were where the biggest issues are.

The other concern was that the project plan relied on assessments that may have issues. I4 mentioned that some of the concept inventories on our list may not be validated or reliable: 
\begin{quote}I think at the FCI, and I think of validation and reliability arguments and there's just none, right? Like, when they developed it, they just you know, I think they did a pretty amazing job for like what they were doing at the time, but if you compare that to the work even done on the BEMA, or at least the published work done on the BEMA, when it was developed, I thought that was like a much more thorough and well-validated instrument.
\end{quote}
I4 argued that better designed instruments are much newer, so the initial time period we would be studying would inevitably consist of manuscripts that contain the issues in our plan. While validity and reliability are fair concerns if we were doing a meta-analysis to cull together findings, our interests are how researchers reported on data and used data to make claims.

\subsection{Advice from the focus group and interviews}
Our focus group participants and interviewees provided a wide-range of advice to improve quantitive research in PER. This includes:
\begin{itemize}
\item \textbf{PER should support community-built software resources and tools.} he statistical programming language, R, was mentioned by 3 individuals as the quantitative analysis tool PER should use. They were enthusiastic that R would be ideal tool to use. They also suggested that the R code used for analysis be shared in some fashion, either in repository or in an appendix in the manuscript. I5 thought this would not only be a means of checking for accuracy but also as a learning tool, suggesting that ``maybe [submitting] commented snippets, so people would understand why did that person do it that particular way.'' 

The authors of this manuscript do not necessarily advocate specifically for R, but they believe that spirit of the sentiment, community-supported tools and sharing code or resources, is worthwhile for PER to consider. We anticipate a broader conversation regarding suitability of tools for different datasets, may be useful for PER.

\item \textbf{Provide enough information that others can check the work.} As readers and reviewers, two individuals were interested in having enough information to determine whether the statistical test was used correctly. I6 has observed instances where they cannot tell whether the researcher was able to use parametric statistics without violating assumptions. F3 said that they sometimes check a manuscript's statistics and advocated for enough information that others can do so: 
\begin{quote}
As a reviewer, occasionally I'll do by hand someone else's t-test to make sure that it comes out. I at least eyeball it... So occasionally, I'll calculate the F to see like "oh, they're saying this is F, but am I nuts, or is it for those mean squares, not seem like it works." ... I think at a minimum, that sufficient statistics to do that has to be reported. That's been APA that have been demanding that for years in order for anything to be published.
\end{quote}
 F3 thought this information does not necessarily need to be in the manuscript's body but could be in an appendix.

\item \textbf{Consider sample size with study design.} I3 noted that early publications in education used small sample, such a single classroom in one school, and later found out that their findings were not applicable. They noted that this could happen to PER. I3's advice regarding sample size and study design is to do one of the following: 

\begin{quote} [My father] goes, ``You know, what's interesting is that within crop science they have a very sort of simple rule about how to approach research studies, and the rule is this: multiple sites, single treatment; single site, multiple treatments.'' That's how that works. If you want to produce a research study that you want to do a single treatment with, then you need to do that research study in multiple sites. The other part of that would be if you want to do a single site, that means you want to go to a single school and do your study, then you need to do multiple treatments. That means you need to do various classrooms, and if there's only one or two classrooms in that school, then you need to do the same treatment in successive academic years.
\end{quote}
\end{itemize}

However, undertaking any of these suggestions might be challenging, particularly if it involves sharing code and providing enough data that one's work can be checked. Researchers may feel vulnerable, as  I1 suggested:
  \begin{quote}
 Yeah, I mean I think it's cultural inertia primarily, right? You know, learning something new is hard and scary. It's also, I honestly think that people are a little scared sometimes to be wrong, right? And so if you put all of your dirty laundry out there, then people will find the dirt. And so that's a little scary for people - it's scary in a world where you think that you will be judged and then not respected for having made a mistake as if none of us have ever made a mistake before. 
\end{quote}
There might be a need to have a cultural change regarding research and our relationship to the tentativeness of science and being wrong, if PER is to adopt any of the advised practices. 

\section{Phase 2: analyzing peer-reviewed manuscripts}
\subsection{Research questions}
Using the data from Phase 1, we designed phase 2 to study peer-reviewed manuscripts that use the following assessments: the Force Concept Inventory (FCI), Conceptual Survey of Electricity and Magnetism (CSEM), Brief Electricity and Magnetism Assessment (BEMA), and Colorado Learning Attitudes about Science Survey (CLASS). Although quantitative PER encompasses more than these assessments or assessments period, we opted to focus on these because they are commonly used by many different researchers. Thus, we can look at an important part of PER that reflects much of the research community. Our research questions are as follows:
\begin{enumerate}
\item  What information on samples have authors reported in peer-reviewed manuscripts that use FCI, CSEM, BEMA, and CLASS?
\item Have authors reported on limitations in these manuscripts?
\item Are claims in these manuscripts supported by data?
\item Do questions 1-3 differ by time period in PER?
\end{enumerate}
Based on the focus group's comments, our underlying hypothesis is that the more recent research tends to report on the sample in more detail, includes limitations, and make claims that are supported by data. 

We note these issues also have been noted in other work (e.g., \citep{Docktor}). The work most similar to ours is a recent paper by Kanim and Cid (2017) that asks which students are studied in PER and whether this is a representative sample in relation to population demographics. While they also examined manuscripts within PER journals, we note that their manuscript and ours differ on several key points. Namely, we looked for presence of information such as a sample description, as the focus group and interviewees suggested quantitative work in PER may lack these descriptions. We also focused on a narrower set of papers; they drew upon papers form the 1970s through 2015 and included a wide variety of papers \cite{Kanim}, while we focused on specific assessments from the 1990s through 2017. Thus, our papers may have similarities but are distinct in their goals and methods.

\subsection{Methodology}
\subsubsection{Journal selection}
For this study, we used manuscripts from AJP, TPT, and PRPER. For ease of discussion, we are using PRPER to refer to both Physical Review- PER and Physical Review Special Topics- PER (PRSTPER), the original name for PRPER. We considered JRST but found few articles that used the assessments of interest.

We selected these journals because they are intended for researchers in PER and they are peer-reviewed. Our interest in peer-reviewed journals is because this study was to examine PER's quantitative work and peer reviewers are part of the research system. Although authors bear the responsibility of creating high quality manuscripts, editors and peer reviewers share this responsibility by ensuring the manuscripts are suitable for publication.

Although PRPER was established because of an identified need for a journal devoted to researchers in PER \cite{Henderson, Singer, Beichner} AJP and TPT historically published important manuscripts. For example, the FCI, the oldest assessment we considered, was introduced in 1992 in TPT \cite{FCI}. The FMCE was introduced in 1998 in AJP \cite{FMCE}. Both instruments are still used today by researchers. AJP and TPT are not opposed to research but emphasize that presented research must have practical applications due to their readership being practitioners.

\subsubsection{Manuscript and assessment selection}
We initially envisioned this project drawing upon manuscripts that use the following assessments:  FCI, Force and Motion Conceptual Evaluation (FMCE), CLASS, Colorado Learning Attitudes about Science Survey for Experimental Physics (ECLASS), Maryland Physics Expectations Survey (MPEX), CSEM, and BEMA. These seven were selected to cover a range of assessments. We also believed these were frequently used. Frequency of use was important to ensure our sample contained enough manuscripts for our study.  Below, we briefly describe each instrument, appearing in chronological order by the first publication on the instrument:
\begin{itemize}
\item  \textit{FCI}, 1992. A multiple choice assessment  designed to measure how well students conceptually understand Newtonian mechanics that are covered in introductory physics courses \cite{FCI}. 
\item  \textit{FMCE}, 1998. A multiple choice assessment designed to measure student understanding of Newtonian mechanics in introductory physics \cite{FMCE}. The FCI and FMCE cover topics to slightly different extents, and the FCI uses pictorial representations for some questions while the FMCE uses graphs for some questions\cite{FMCE2}. 
\item \textit{MPEX}, 1998. A multiple choice assessment designed to measure student beliefs and attitudes towards  physics, including how they approach studying in the course and  whether they think physics is relevant to them \cite{MPEX}. Respondents select how much they agree with a statement on five-point Likert scale (strongly disagree to strongly agree) \cite{MPEX} 
\item \textit{CSEM}, 2001. A multiple choice assessment designed to measure student understanding of introductory electricity and magnetism \cite{CSEM}. The authors use both graphical and pictorial representations in this concept inventory \cite{CSEM}.
\item  \textit{CLASS}, 2006. Designed to study students' beliefs about physics and learning physics, such as whether learning physics has any usefulness to them in their lives  \cite{CLASS}. Respondents rate statements on a five-point Likert scale (strongly disagree to strongly agree) \cite{CLASS}.  
\item \textit{BEMA}, 2006. A multiple choice assessment designed to measure student understanding in traditional calculus-based electricity and magnetism (E\&M) as well as those enrolled in E\&M courses that use \textit{Matter and Interactions II: Electric and Magnetic Interactions} curriculum \cite{BEMA}. Similar to the other concept inventories on this list, the emphasis is on conceptual understanding rather than mathematical calculations \cite{BEMA}.
\item \textit{ECLASS}, 2012. Designed to study changes in student attitudes towards laboratory practices before and after a lab course \cite{ECLASS}. ECLASS uses a five-point Likert scale (strongly disagree to strongly agree). Items on this assessment ask respondents to consider how they feel about different statements and asks respondents how a hypothetical physicist might answer \cite{ECLASS2}.
\end{itemize}
We looked at each assessment on PhysPort, a website that, among other services, has a comprehensive collection of assessments \cite{PhysPort}. Only ECLASS has built-in demographic questions, both for the individual student respondent as well as course-level information (e.g., institution name, course name). Because ECLASS had built-in demographic questions and was fairly recent, we decided to exclude manuscripts that used it for uniformity. On PhysPort, each assessment has an implementation guide that describes the purpose of each assessment. The assessments link to a best practices guide that mentions studies that have examined demographic differences (e.g., gender, race/ethnicity)  but does not suggest that those data are collected \cite{PhysPort2}. 

To ensure the number of manuscripts studies was manageable, we decided to include manuscripts from the first five years of the instrument's introduction and the most recent five years to reflect contemporary practice. For example, we included manuscripts that used the FCI that were published from 1992 to 1997 as well as manuscripts from 2012 through 2017.  We excluded any meta-analyses, because their reporting capabilities are reliant on the original studies, as well as studies that only mentioned the assessment in their literature review.

Our searches in the journals were based on the history of publishing in PER. We searched for manuscripts in AJP and TPT if the instrument was introduced prior to 2010. Although the PRPER began in 2005, fewer than 10 articles were published in 2005 and approximately 25 were published in 2009 \cite{Beichner}; we suspected that researchers may not have initially been aware of PRPER or still saw AJP or TPT as the best options to publish PER work. If the instrument's first five years coincided with 2005, the debut of PRPER, we searched in PRPER as well. For example, we looked for manuscripts that used CSEM (introduced in 2001 \cite{CSEM})  in AJP, TPT, and PRPER because the first five years ranged from 2001-2006.  We only looked in PRPER for recent papers, the 2012-2017 timespan, because three hundred eighty-eight papers had been published \cite{PRPERSearch1}. This suggested that PRPER would provide an adequate number of articles to study in the more recent timespan.

Manuscripts were further examined by Aiken and Knaub to determine to what extent the assessment was used. Our list was further refined to only include manuscripts that focused on the assessment data; some manuscripts used assessments in the periphery, only discussing them in a few sentences and focusing on other research areas. We found 84 manuscripts that met our criteria. Thirteen used more than one of the assessments of interest.

\begin{table*}
\label{AssessmentBreakdown}
\caption{Description of assessments considered for Phase 2}
\begin{tabular}{ll|lllllllllll}
\textbf{Assessment} &&\textbf{Range of}  &&& \textbf{No. of articles}  &&& \textbf{No. of articles} &&& \textbf{Total no. of articles}\\
&&\textbf{first five years} &&&\textbf{from first 5 years} &&& \textbf{from most recent 5 years} &&&  \textbf{using assessment}\\
\hline 
FCI&& 1992-1997 &&& 3 &&& 29 &&& 32\\
FMCE && 1998-2003 &&& 2 &&& 4 &&& 6\\
MPEX&& 1998-2003 &&& 3 &&& 5 &&& 8\\
CSEM&& 2001-2006 &&& 5 &&& 9 &&& 14\\
BEMA&& 2006-2011 &&& 8 &&& 4 &&& 12\\
CLASS&& 2006-2011 &&& 11 &&& 15 &&& 26\\
\end{tabular}
\end{table*}

 Table I displays the range of the first five years of each instrument, as well as the number of articles we found for the first five years and the most recent 5 years for the FCI, FMCE, MPEX, CSEM, BEMA, and CLASS. We note that the total is greater than 84 because some manuscripts used more than one assessment. We only counted the use of an assessment if it coincided with the first five years of the assessment's existence. For example, a manuscript published in the early 2000s might use both the FCI and CSEM. We would only count this as a CSEM paper, because the FCI would have existed for a decade.

We included only assessments that appeared in 10 or more manuscripts, ending with FCI, CSEM, BEMA, and CLASS. Four included manuscripts used FMCE or MPEX in addition to one of the four assessments included in this study. This left us with 72 manuscripts. Seven are from AJP, sixty-two are from PRPER, and three are from the TPT. These manuscripts have 176 researchers as lead or co-authors. Twenty-four (33.3\%) of these manuscripts studied non-US populations (e.g., students from Canada, students from China). The list of manuscripts we coded can be found online \cite{Github}.

\subsubsection{Manuscript coding scheme and methods}
A priori codes were developed based on the focus group, the interviews, and the researchers' experiences with reporting sample descriptions, limitations, and conclusions. Codes delved into population context descriptions, sample descriptions, discussion of limitations, and how findings were used to support conclusions. Coding was done to look for the presence of these features, not making judgments about how well these features were described.

Aiken and Knaub initially coded 15 manuscripts independently, reading each manuscript to find the particular information. They used the a priori codes as well as noted any emergent codes. They met to discuss emergent codes, difficulties in interpreting the codes, and discrepancies between their coding. The coding scheme was modified. Using the new coding scheme, they independently coded all manuscripts including the 15 they previously coded. Their coding was then combined. When discrepancies in coding occurred, the manuscripts were rechecked. 

Table II displays the final coding scheme. Coding erred on the side that the information was present, even if the information was difficult to find. For example, a manuscript by authors at one institution might say ``our students'' when discussing the data and make references that they were the instructors, but not name the institution. This would be coded as the institution was named, even though it required more careful reading than manuscripts that named the institution in the body. We also read similarly for the codes under limitations and conclusions.

\begin{table*}[t!]
\label{Coding Scheme}
\caption{Coding scheme for Phase 2}
\begin{tabular}{ll|lllllllllllll}
\textbf{Category} && \textbf{Code}  &&& \textbf{Description} \\
\hline 
\hline
\textbf{Institution/course context}&&\textit{Location} &&& Where the institution is located. Includes general  \\
&&&&& (e.g., Midwest) and specific (e.g., Boston, MA) \\
&&\textit{Institution name} &&& The name of the institution (e.g., Michigan State)\\
&&\textit{Population demographics}&&& Any descriptions of the population studied (e.g., gender \\
 &&&&& of students in the course, percentage of international \\
 &&&&& students in the institution, etc.)\\
&&\textit{Course description}&&& Any description of the course (e.g., taught using active \\
 &&&&& learning, introductory physics)\\
 \hline
\textbf{Sample} &&\textit{N}&&& The authors reported the total number of responses\\
&&\textit{Response rate}&&&The authors reported how many responses they received\\
&&&&&  relative to the number of potential respondents. We   \\  
 &&&&&  counted both manuscripts that explicitly had a response  \\
 &&&&& rate and those that provided enough data to find one (e.g.,  \\
  &&&&& including both the number of  responses and the total  \\
  &&&&& number of students in a class)\\
&&\textit{Sample demographics}&&& Any description of the sample (e.g., race, gender, majors)\\
&&\textit{Background in physics}&&& Any description of the sample's prior experiences in \\
  &&&&& physics (e.g., participants had taken at least one physics\\
  &&&&& course)\\
&&\textit{Course grade}&&&Any description of the course grades of the students in the\\  &&&&&  sample (e.g., letter grades, numeric grades)\\
&&\textit{Year in schooling}&&& Any description of year in schoolings of the sample (e.g., \\
  &&&&&sophomores, recent graduates)\\
&&\textit{Instructor/section}&&& Any description regarding who taught the course (e.g.,      \\
  &&&&& teaching background of the instructor, number of  \\
  &&&&&  instructors) or how many sections are in the sample \\
   \hline
\textbf{Limitations}&&\textit{Sample and population limitations} &&& Any acknowledgement that the sample and/or population  \\   &&&&&may impact the results such that generalizability or    \\
  &&&&&  applicability may hindered (e.g.,  collecting from one  \\
  &&&&& institutions, the sample only has students who passed the \\
  &&&&& course)\\
&&\textit{Statistical and study design limitations}&&&  Any acknowledgement that the statistics and/or the study  \\
  &&&&&  design impacts the results such that the results such that \\
     &&&&& generalizability or applicability may hindered (e.g., causal \\
     &&&&&  claims cannot be made, noting lack of a comparison group)\\
&&\textit{Attempt to overcome limitations}&&&  The use of any technique done to mitigate a limitation (e.g., \\
&&&&&   using paired data or imputation methods for incomplete  \\
&&&&& data sets) \\
 \hline
\textbf{Conclusions}&&\textit{Data are used to support conclusions}&&&  Conclusions are made referring to the data \\
&&\textit{Conclusions do not overgeneralize or overstate} &&& Conclusions acknowledge that the results may not be\\
&&&&&  universal (e.g., conclusions acknowledge limitations,   \\
&&&&& concluding statements refer to the study and the sample)  \\
&&&&& and present claims tentatively (e.g., authors do not make \\
&&&&& absolute statements) \\
\hline
\hline
\end{tabular}
\end{table*}

The codes reflect the potential issues in quantitative PER work our focus group and interviewees noted. We note that some codes are more specific (e.g., \textit{Institution name}) than others (e.g., \textit{Population demographics}). We were more specific for some codes because the information would more readily be available and standard, such as the \textit{Institution name}, \textit{Course grade}, and \textit{N}. The less specific codes were a compromise of noting that authors provided some description but not applying standards that may be harmful to their sample (e.g., gender information may inadvertently reveal respondents), an issue noted in Phase 1 of this study.

\subsubsection{Limitations}
Phase 2 has several limitations regarding the scope of this overall project. The primary limitation is that we are focused on a handful of assessments. They do not encompass the entire body of quantitative research in PER. Perhaps non-assessment quantitative work or different assessments would yield different results. Despite this limitation, focusing on these assessments had some advantages. Pragmatically, we were able to set a boundary around a manageable set of manuscripts and could thoroughly examine them. Although these assessments focus on different aspects of physics education, they are similar in that they do not have built-in demographics questions. This eliminated some variability. 

A second major limitation is that we drew upon three distinct journals who have different purposes. The target audiences are different, so the inclusion or exclusion of particular information (e.g., limitations) may not feel as relevant if a manuscript is for practitioners. However, PRPER only came into existence in recent times. Drawing upon AJP and TPT are best options to determine whether PER has improved within our scope. 

Lastly, we coded manuscripts for the existence of these particular features. We cannot make any claims to the quality of sample reporting or whether manuscripts did an adequate job of addressing limitations and discussing conclusions. Given the wide range of contextual situations regarding samples and manuscript research questions and goals, we felt that looking for the existence of these features was adequate for the goals of this manuscript. We suggest that the quality of these manuscript features be future work for interested researchers.

\begin{figure*}[ht!]
\label{Fig1}
\includegraphics[scale= 0.4]{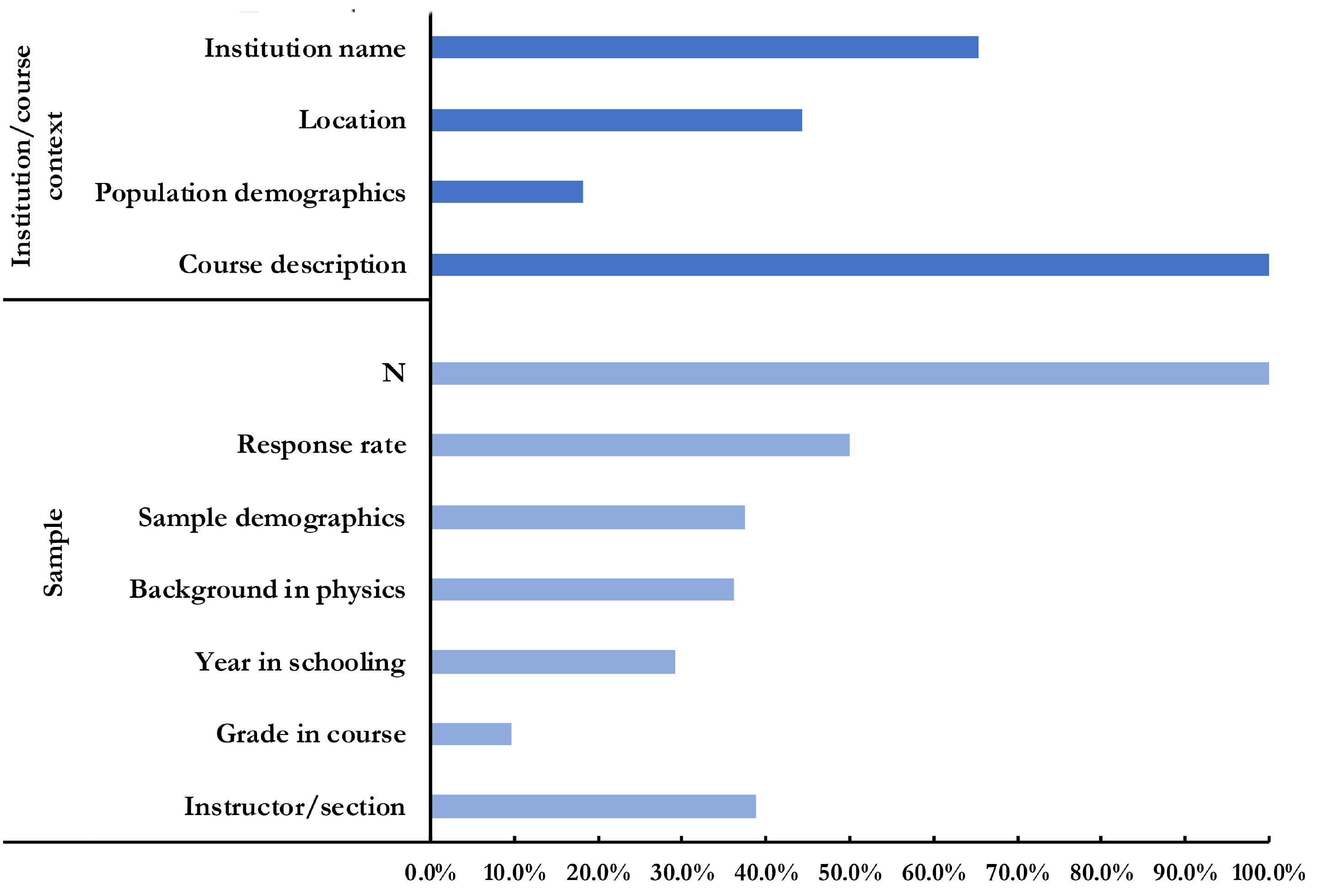}
\caption{Percentage of manuscripts that had specific population and sample codes}
\end{figure*}
\subsection{Results: information reported on institutional/course context and sample}

Figure 1 displays the percentage of manuscripts that were found to contain different features. The only features that all manuscripts in this sampled contained was \textit{N} and the course description. Many (47 or 65.3\%) reported the institution name and half of the manuscripts reported a response rate. It is unclear why response rate was not reported. The manuscripts in this study had well-defined samples (e.g., a particular course or several courses), suggesting that the authors knew the total number of possible respondents.

\subsubsection{Time period in PER}
\label{SampleTime}
The data were further divided as ``early'' and ``recent.'' ``Early'' were manuscripts published before 2012, while ``recent'' were manuscripts published from 2012-2017 (the most recent completed 5 years). Twenty-five manuscripts are in the ``early'' category. Forty-seven are in the ``recent'' category. 

Figure 2 displays these data. All data are presented as percentages of ``early'' or`` recent.'' The features show an assortment of differences. Some features such as \textit{Response rate} show an increase between ``early'' and ``recent'' while other codes show a decrease, such as \textit{Population demographics} and \textit{Instructor/section}. \textit{Institution name}, \textit{Year in schooling}, and \textit{Sample demographics} had similar percentages of reporting in both ``early'' and ``recent.''

\begin{figure*}[ht!]
\label{Fig2}
\includegraphics[scale= 0.4]{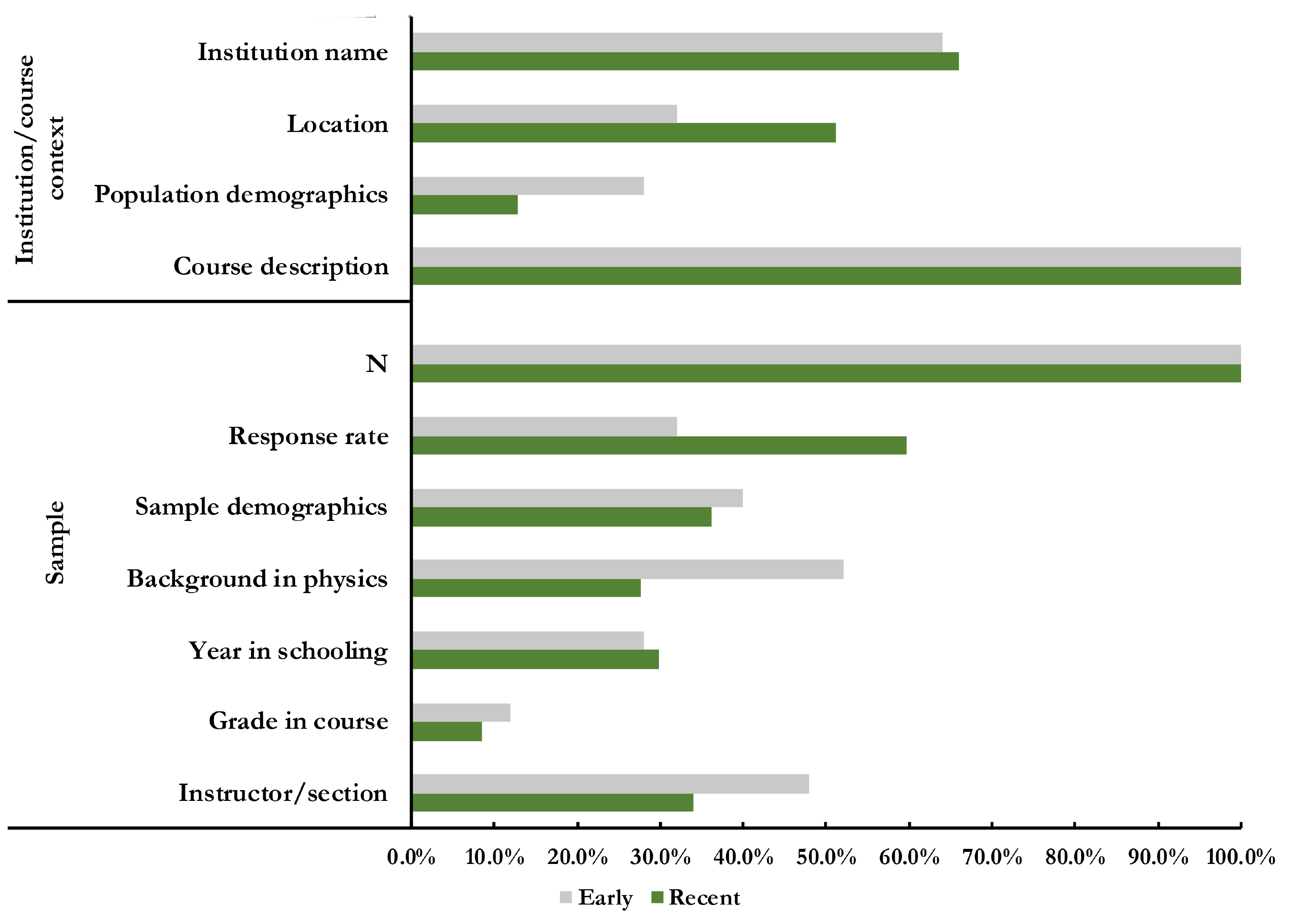}
\caption{Percentage of manuscripts that had specific population and sample codes, disaggregated by recent (2012-2017) and early (1992-2011)}
\end{figure*}

We ran Fisher's Exact test on each feature to determine whether there is any statistically significant relationship between each feature and the early/recent category. Fisher's Exact test is similar to the chi-square ($\chi$ \textsuperscript{2}) test and is generally preferred for small samples sizes. Table III displays the $\chi$ \textsuperscript{2} results, the p-value for Fisher's exact test, and the effect size ($\phi$). Because \textit{Course description} and \textit{N} were reported on by all manuscripts, we did not include them in the table.

\begin{table}[h!]
\caption{Fisher's Exact test results on institutional/course and sample features}
\begin{tabular}{ll|lllllllllllllllll}
\textbf{Feature} &&\textbf{$\chi$ \textsuperscript{2}} &&&& \textbf{P-value}&&&& \textbf{$\phi$}\\
\hline
\textit{Institution name} && 0.028 &&&& 1.000 &&&& 0.020\\
\textit{Location} && 2.402 &&&& 0.142 &&&& 0.183\\
\textit{Population demographics} && 2.506 &&&& 0.123 &&&& -0.189\\
\textit{Response rate}&& 4.963 &&&& 0.047\textasteriskcentered &&&& 0.263\\
\textit{Sample demographics}&& 0.102 &&&& 0.801 &&&& -0.038\\
\textit{Background in physics}&& 4.191 &&&& 0.07 &&&& -0.241\\
\textit{Year in schooling}&& 0.025 &&&& 1.000 &&&& 0.019\\
\textit{Grade in course} && 0.226 &&&& 0.688 &&&& -0.056\\
\textit{Instructor/section}&& 1.338 &&&& 0.312 &&&& -0.136\\
\hline
\end{tabular}
\\

\textasteriskcentered = \textit{p} < 0.05\\
\end{table}

Almost all features have small effect sizes (<$\mid$0.2$\mid$, per traditional convention)and were not statistically significant. In other words, manuscripts in the ``early'' period are not very different from manuscripts in the ``recent'' period, on a whole, for the majority of features for the institutional/course context and sample categories.

The only feature that has a statistically significant (\textit{p} < 0.05) result is \textit{Response rate}, which suggests a relationship between this feature and the early/recent variable (i.e., results are less likely to chance).  Though the percentage of manuscripts with  a \textit{Response rate} almost doubled in the ``recent'' times, the effect size is somewhat moderate (> 0.2). This means there is a modest difference between early and recent periods.

\subsection{Results: reported limitations}
Figure 3 displays the percentage of manuscripts that reported any limitations, reported sample limitations, reported statistical or study design limitations, and attempted to address or overcome the limitations. We note that there may be no good way to overcome limitations, but we were interested because one of our interviewees suggested that authors do not often attempt to overcome study limitations.

\begin{figure*}[ht!]
\label{Fig3}
\includegraphics[scale= 0.4]{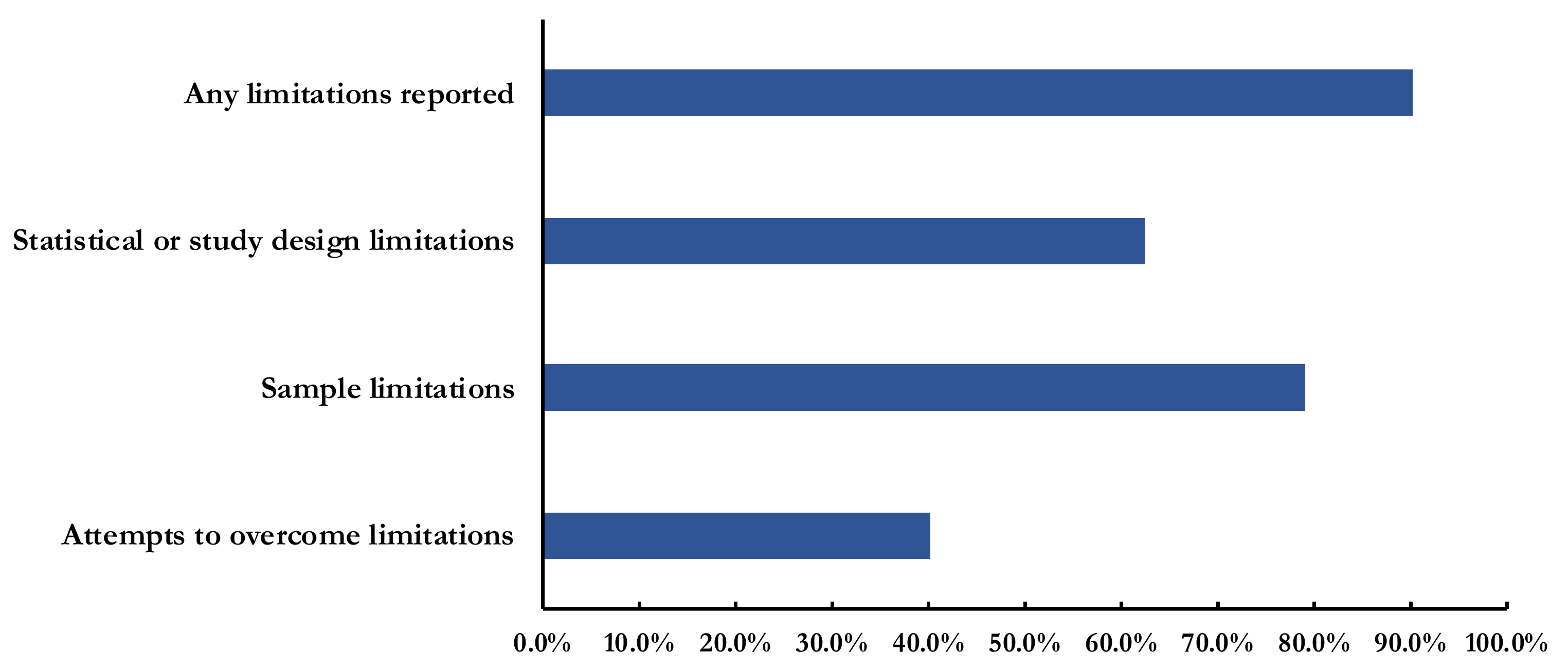}
\caption{Percentage of manuscripts that had reported on limitations}
\end{figure*}

While most manuscripts (approximately 90\%) in this study reported limitations, a few did not. Slightly over half (37 or 51.4\%) reported both on sample and statistical or study design limitations. Fewer than half of the manuscripts noted some kind of attempt to overcome the study's limitations.

\subsubsection{Time period in PER}
The data were analyzed to determine whether manuscripts written in earlier times were different from those written more recently. Similar to section \ref{SampleTime}, we ran Fisher's exact test to see if differences were statistically significant. These results are displayed in Figure 4 and in Table IV.
 
All manuscripts in recent times note at least one limitation. Reporting limitations on samples, limitations in statistics or study design, and attempts to overcome limitations are have all increased from earlier to recent times. These are all found to be statistically significant and have moderate effect sizes. This suggests these results may not be due to chance and that the difference is somewhat substantial.

\begin{figure*}[t!]
\label{Fig4}
\includegraphics[scale= 0.4]{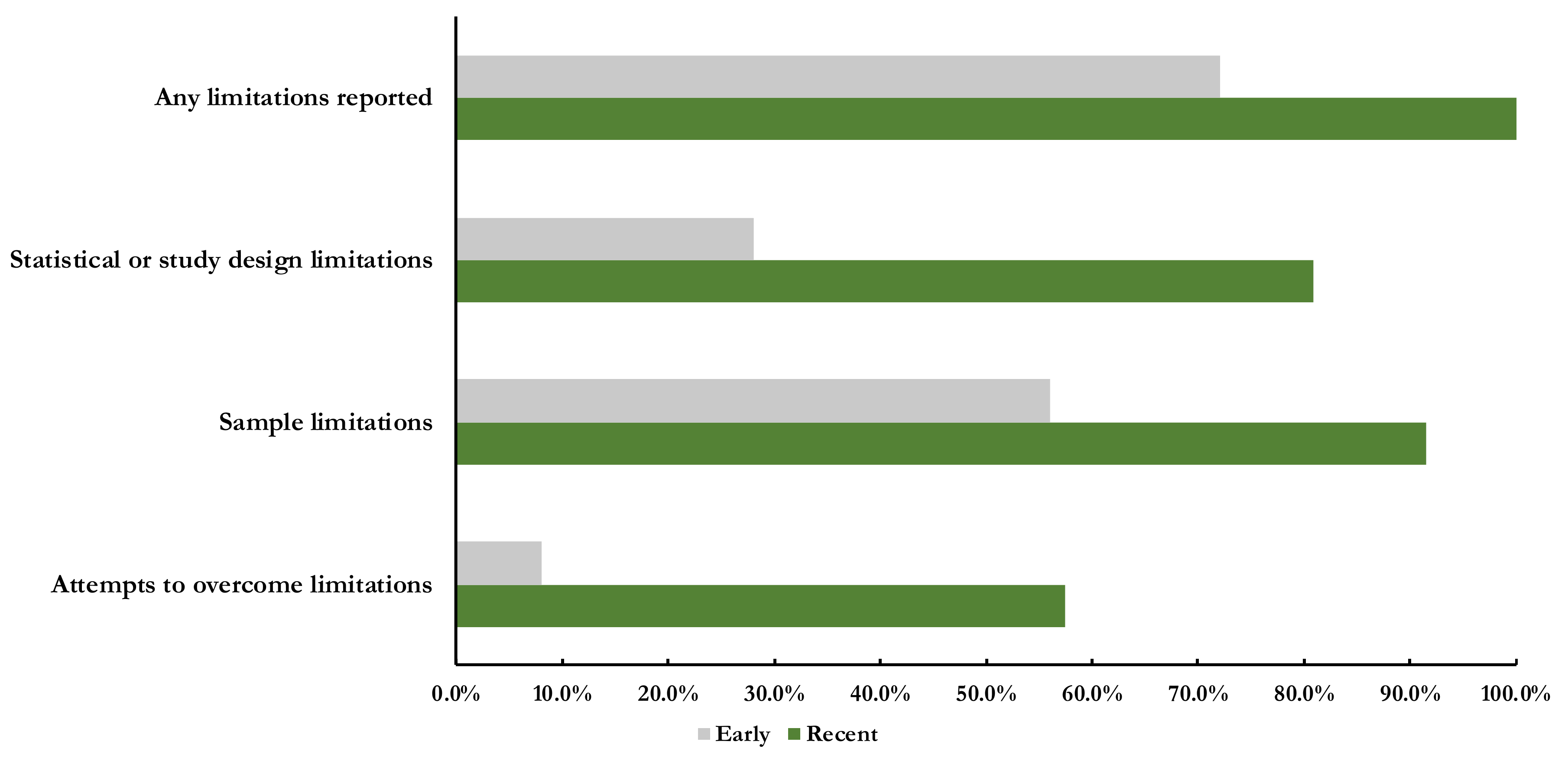}
\caption{Percentage of manuscripts that had reported on limitations, disaggregated by recent (2012-2017) and early (1992-2011)}
\end{figure*}

\begin{table}[h!]
\caption{Fisher's Exact tests on limitation features}

\begin{tabular}{ll|lllllllllllllllll}
\textbf{Feature} &&\textbf{$\chi$ \textsuperscript{2}} &&&& \textbf{P-value}&&&& \textbf{$\phi$}\\
\hline
\textit{Any limitations reported} && 16.6 &&&& 0.000\textasteriskcentered &&&& 0.48\\
\textit{Statistical or study design limitations} && 19.4 &&&& 0.000 \textasteriskcentered&&&& 0.52\\
\textit{Sample limitations} && 12.5 &&&& 0.001\textasteriskcentered &&&& 0.42\\
\textit{Attempts to overcome limitations}&& 16.6 &&&& 0.000\textasteriskcentered &&&& 0.48\\

\hline
\end{tabular}
\\
\textasteriskcentered = \textit{p} < 0.05\\
\end{table}

\subsection{Results: reporting on conclusions}
Lastly, we looked at whether the conclusions referred to the data in the study and whether the conclusions overgeneralized or overstated findings. These findings are in Figure 5. All manuscripts in this study made use of the study's data to support any conclusions. Most manuscripts did not overgeneralize or overstate their conclusions. 

\begin{figure*}[t!]
\includegraphics[scale= 0.37]{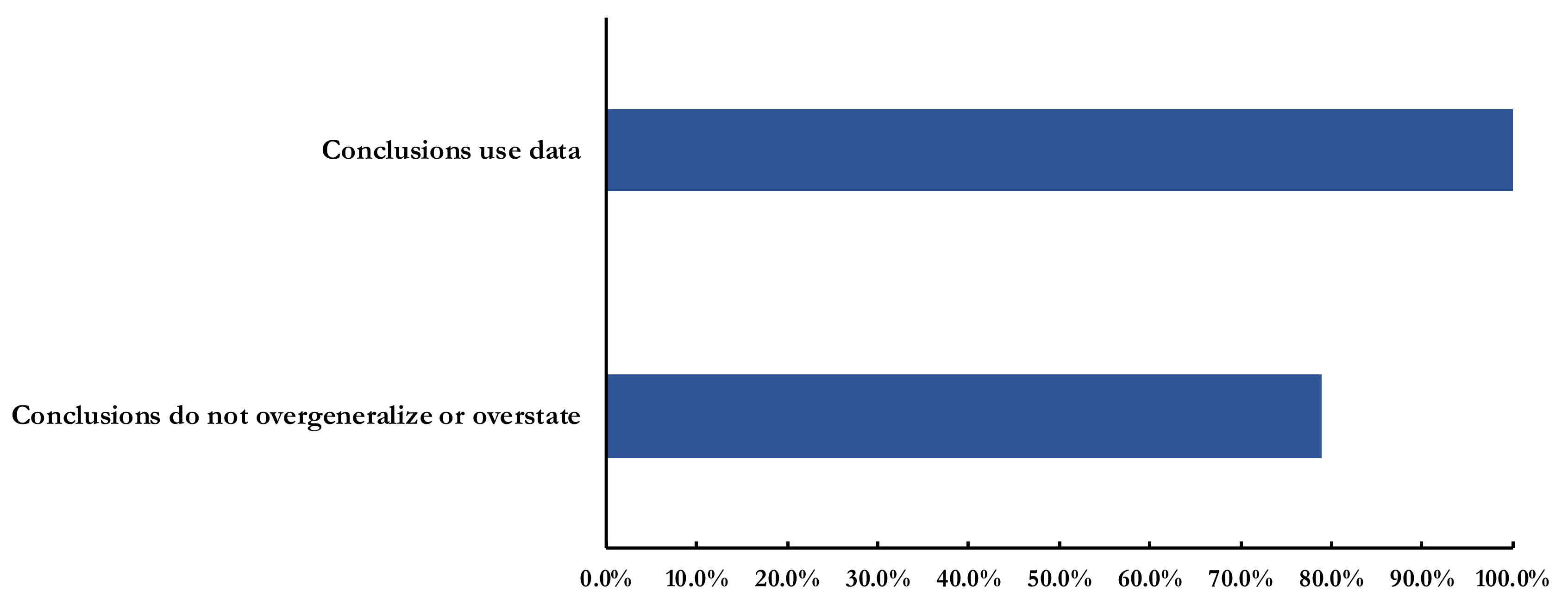}
\caption{Percentage of manuscripts that used data in the conclusions and conclusions did not overgeneralize or overstate}
\end{figure*}

\subsubsection{Time period in PER}
We examined the issue of overgeneralization to see if recent manuscripts in this study differed from earlier manuscripts. These results are in Figure 6. Manuscripts in more recent times tended not overgeneralize or exaggerate their conclusions, though some (12.8\%) still do. This result is statistically significant, though the effect is somewhat small ($\chi$ \textsuperscript{2} = 5.34, \textit{p} = 0.032, $\phi$ = 0.272). This suggests that these results may not have been due to chance and that the difference is small.

\begin{figure*}[t!]
\includegraphics[scale= 0.37]{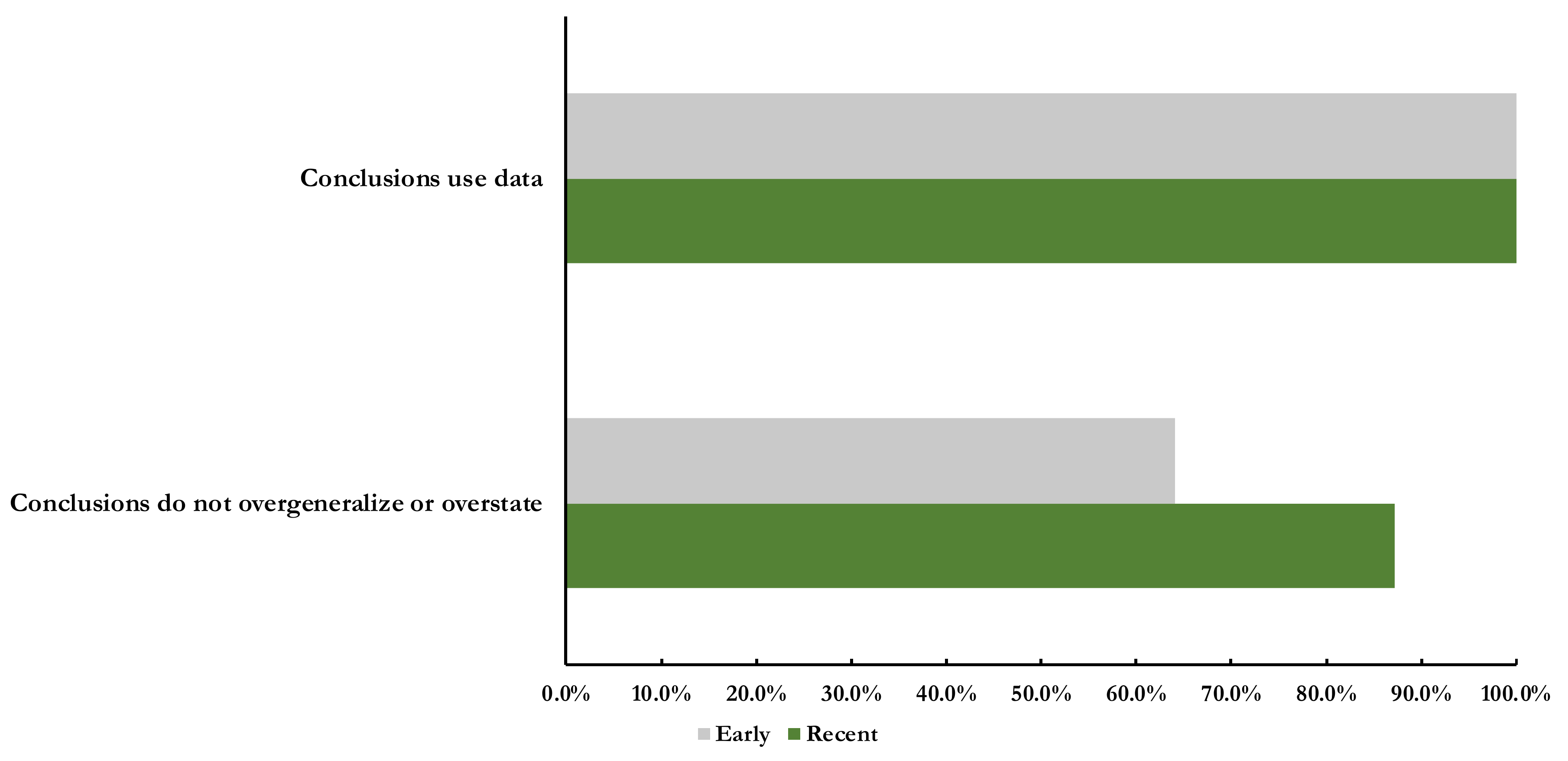}
\caption{Percentage of manuscripts that used data in the conclusions and conclusions did not overgeneralize or overstate, disaggregated by recent (2012-2017) and early (1992-2011)}
\end{figure*}

\section{Discussion and conclusions}
\subsection{Phase 1 discussion}
This two-phase study's goals were to find out what are pressing issues in quantitative PER and to determine how pervasive these issues are. During Phase 1, we ran a focus group of experts in quantitative PER who were identified by editorial members of PER publications. The focus group's main concerns were manuscripts having poor sample descriptions, not reporting limitations well, and overgeneralizing/overstating conclusions. They pointed out that much of the quantitative work in PER is focused on assessments. The focus group also believed that these issues were resolving themselves as PER matured. 

The results from this phase were particularly interesting because the issues mentioned are fundamental aspects of research, regardless of whether it is quantitative. Although these issues are fundamental aspects of research, resolving these issues may be complicated and non-prescriptive. As our focus group and interviewees noted, there are legitimate reasons to not report overly detailed descriptions on samples. Limitations and generalizability will vary from study to study. Some of this may depend on the author's intended audience; for example, authors who write for practitioners may fear that too much space devoted to discussing limitations will deter practitioners from reading their manuscript. Still, there is some general sense that manuscripts should include these aspects.

\subsection{Phase 2 discussion}
After the focus group, we created a project plan and interviewed additional experts to provide feedback. The interviewees were mostly in agreement with the plan, though they did express some caution in applying any universal standard to some reporting on these issues (e.g., demographic data on the sample). This sentiment was also expressed by the focus group. During Phase 2, we looked at manuscripts that use the FCI, CSEM, BEMA, and CLASS from AJP, PRPER, and TPT, all peer-reviewed PER journals. We were interested in what is reported on samples and limitations as well as how conclusions are articulated. We were also interested in whether there were any difference between early (1992-2011) publication times of these assessments and recent times (2012-2017). Our general hypothesis, based on the focus group, was that these manuscript features are more common in recent times.

The manuscripts in our study have used a variety of ways to describe the institutional/course context and the sample in their studies. Although there are differences between early and recent times, most were not  found to be statistically significant via Fisher's Exact test. This suggests that reporting on these features has not changed much since the early days of these assessments and present time. One feature that was found to be statistically significant,  \textit{Response rate}, had small effect size. We note that the four assessments in this study do not have built-in questions regarding demographic information or institutional/course context. Future work is needed to see whether built-in questions have any effect on sample or institutional/course context descriptions. 

We urge caution in judging manuscripts that report these features and in turn, comparisons between ``early'' and ``recent'', as ``good'' or ``bad.'' We note that a third of the manuscripts studied individuals from non-US institutions; there may be different legal and ethical standards for reporting. Additionally,  some features, such as \textit{Location} or \textit{Institution}, may not have a readily apparent purpose, making a judgment of ``good'' difficult. For example, we are unsure what is gained by knowing that the students were from the East Coast. Even knowing an institution name sometimes does not help if one is not familiar with the institution. Perhaps there is, as described in Phase 1, an implicit theory that compels authors to include this information. We do not bring this issue up to shame authors who have done so; we, the authors of this manuscript, have also reported some of these features without much if any explanation. 

Still, while we respect that not all of these features can be included due to ethical obligations to research participants, we are unsure why some features are not reported. Some features should not violate ethical obligations and would be useful. For example, while reporting response rates has increased in more recent times, a little under 60\% of the recent manuscripts in our study report a response rate. These studies had bounded systems (e.g., a course), meaning a total number of possible respondents is known. We also noticed that few studies indicated whether their sample was representative of the population of interest (e.g., were students with A's overrepresented in the sample?). These features can help readers understand the limitations of the study and perhaps help researchers gain additional insights on their work.

The recent manuscripts in this study have more reporting on limitations and have more often mentioned ways in which the authors attempted to overcome limitations. Recent manuscripts are also less likely to overgeneralize or overstate in the conclusions. While these results are encouraging, we ponder whether most of these features should be present in all manuscripts. At the very least, no manuscript should overgeneralize or overstate their conclusions.

\subsection{Conclusions}
Within the context of this study, manuscripts that use the FCI, CSEM, BEMA, and CLASS have improved since the early days in that limitations are present and that conclusions are are not overgeneralizing. Sample reporting has not changed much, though we are cautious to suggest whether that is overall negative.

We emphasize that these results are limited to the context of this study and that readers should not interpret these results to mean that reporting on samples, limitations, and conclusions is fine. As noted earlier, we erred on the side of these features being present even if they were subtly indicated (e.g., writing that discusses ``our students'' counted as not overgeneralizing, refer to another paper that describes the study in better detail). Aiken and Knaub compared coding, and the manuscripts were re-examined when discrepancies occurred. These were quickly resolved, but we note that we read these papers for research purposes and were deliberately looking for these features. A typical reader, who is not conducting such a study, may not read a manuscript in this manner. In short, these features are present but may not be easily found or may be accidentally overlooked.

Similarly, we also did not code for how well manuscripts reported any of these features. We coded for these features being present. There are likely manuscripts in this study that do not cover the most important or relevant sample descriptions or limitations.

Despite these caveats, these results shed some light on quantitative PER. We see these results as a step towards critically examining quantitative work in PER. To aid this work in critically examining quantitative work, we offer the following questions for PER to consider when writing or reviewing manuscripts. These are just questions, not necessarily with a ``correct answer.'' 
\begin{itemize}
\item What information regarding the sample is useful for the audience?
\begin{itemize}
\item What implicit messages could the included information tell the audience?
\item Does the included information need explanation so that the audience understands why it is important?
\item  If the author does not include particular information regarding the sample, is the research weakened as result?
\end{itemize}
\item Are sample descriptions and limitations implicit? Should they be explicit?
\item Which limitations are important for authors to acknowledge?
\begin{itemize}
\item Is it adequate for the author to just acknowledge the limitations?
\end{itemize}
\item How explicit should authors be that their conclusions do not overgeneralize? 
\item How strongly should authors make their claims?
\item How much effort should the audience have to make in finding any of this information?
\begin{itemize}
\item If sample descriptions and limitations are not carefully woven throughout the manuscript, what message might a reader receive?
\end{itemize}
\end{itemize}
The authors of this manuscript acknowledge there is a great deal of nuance to how these questions could be answered, and there likely is no ``one-size-fits-all'' response to any of these questions. However, we do believe that these questions are important for researchers, reviewers, and readers of PER to consider. This responsibility is not just for the researchers who create the studies and write the manuscripts but for all involved in the research enterprise. Manuscripts are peer-reviewed and research is used, cited, or applied. 

As these studies are used to inform practice and policy that can affect many, it is important that studies do not  misinform even if misinformation is inadvertent. Future work is needed to examine the quality of reporting on these issues, as well as work on exploring how readers engage and understand  research. We anticipate such work would help researchers understand how to best communicate their results. PER as a research community may benefit from an open conversation regarding what should be presented and why. Again, we do not have answers but believe that further research and exploring questions like the ones we propose, even in a conversation, could lead to more robust research that ultimately leads to better physics education.

\section{Acknowledgements}
The authors thank the editors and advisory boards of Physical Review- Physics Education Research and the American Journal of Physics for their suggestions for potential interviewees/focus group members. The authors thank the focus group participants and interviewees. Chastity Aiken provided additional feedback and proofreading that was much appreciated. We also thank Michigan State University, the Ohio State University, and the University of Oslo.

This study was partially funded by the Norwegian Agency for Quality Assurance in Education (NOKUT) which supports the Centre for Computing in Science Education at the University of Oslo.

\appendix
\section{Focus group protocol}
{\em Facilitator introduction: Thank you for joining us for this focus group. As we had described in the email invitation, we are interested in your observations and opinions of statistical use in PER. You all were invited  because  you have substantial experience in this area.
\\
\\
The focus group will be recorded and transcribed. While we will de-identify the data of your names and institutions, we may use some quotes in the manuscript we will produce.
\\
\\
If you are okay with being recorded, please say yes. [Pause for the focus group to say yes]. If you did not say yes, we ask that you leave the call. Pause for anyone who wishes to leave. I will now turn on the recorder.}

\begin{enumerate}
\item Please state your name, the institution where you work, and roughly how long you have been participating in PER.
\item When you think about statistical and quantitative work in PER in published work, what comes to mind?
\begin{enumerate}
\item What kind of mistakes do you often see being made?
\item What kind of issues do you often see that make it challenging to determine whether the methods were sound?
\item Which paper(s) do you feel are good examples of quantitative work? What particular aspects make them good examples?
\end{enumerate}
\item We have a list of potential and/or common quantitative and statistical mistakes and misuses from the literature. {\em Shares screen with focus group participants}. Have you often seen these kind of errors before in PER work?
\item Based on your comments, the most prevalent errors described in the group are {\em reads off list off errors}. Did we accurately summarize the discussion? If not, what changes should we make?
\begin{enumerate}
\item {\em If there are a lot of errors} Which mistakes and issues do you feel are the most important for us to focus on? Why?
\end{enumerate}
\item What resources would be useful for the PER community so that they make better use of quantitative methods and statistical analysis?
\begin{enumerate}
\item {\em If the resource already exists} Would you please provide the URL/title/etc. of this resource?
\item {\em If the resource does not exist} Is there an example in another field that is similar to what you have in mind?
\end{enumerate}
\item Is there anything else you feel is important for us to know?
\end{enumerate}
{\em Facilitator closes focus group, thanks participants, and shuts off recorder.} 
\section{Interview protocol}
\begin{enumerate}
\item Would please tell me roughly how long you have been participating in PER and a bit about your research background? {\em Keep this part brief}
\item Recently, we ran a focus group to find out perspectives on quantitative methods in PER. Participants included individuals who were recommended by editorial staff for PR- PER and AJP. We were interested in what they felt were the important issues in quantitative and statistical work in PER in terms of mistakes, ambiguities, and misuses. Based on their feedback, we created the project plan we had sent to you. We would like you to comment on it. 
\\
\\
Specifically, we are interested in:
\begin{enumerate}
\item Your feedback on this topic and whether we should be focused on different topics.
\item Whether the work plan answers important and relevant research questions to quantitative researchers.
\end{enumerate}
\item {\em If interviewee does not think the work plans covers a good topic but has no suggestions of their own} We provided the focus group with an a priori list that contained a list of technical issues in quantitative research, based on issues identified in other social sciences. Which of these issues do you think are most important for us to cover in this project? Can you give me some specific examples or reasons why you think these are the most important? 

\item {\em If interviewee does not think the work plans covers a good topic AND has their own suggestions} Can you give me some specific examples or reasons why you think these are the most important? 

\item Additionally, we are interested in collecting resources for quantitative researchers in PER. What resources (such as publications, programs, repositories, or other resources) would be useful for the PER community so that they make better use of quantitative methods and statistical analysis?
\begin{enumerate}
\item {\em If the resource already exists} Would you please provide the URL/title/etc. of this resource?
\item {\em If the resource does not exist} Is there an example in another field that is similar to what you have in mind?
\item Which paper(s) do you feel are good examples of quantitative work? What particular aspects make them good examples?
\end{enumerate}
\item Is there anything else you feel is important for us to know?
\end{enumerate}
\section{Project plan for interviewees}
{\em (This was a document given to each interviewee to comment on)}

\subsection{Phase 1}
We held a focus group of recommended quantitative researchers who are either quite familiar with PER or researchers in PER. We were interested in finding out what they felt were the primary issues in quantitative PER works. They noted that much of the quantitative work in PER is focused on assessment. Members of the focus group were mostly concerned with the following:
\begin{itemize}
\item The research contains little to no description of the sample (e.g., demographic information and institutional context).
\item Researchers may try to generalize to all students/institutions when they sample only included selective, predominantly white institutions.
\item Claims are not supported by the data.
\end{itemize}
The focus group also emphasized that these issues, among others, are improving. They pointed out that PER is a young field and believe that some of these issues are resolving themselves.

\subsection{Phase 2}
As we had described in our Phys. Rev.- PER manuscript proposal, we plan on looking through peer-reviewed articles to determine how pervasive the issues noted by the focus group are.

Based on their comments, we are considering focusing on the following student assessments:
\begin{enumerate}
\item FCI
\item FMCE
\item CLASS
\item MPEX
\item ECLASS
\item CSEM
\item BEMA
\end{enumerate}

Because the focus group emphasized that quantitative work is improving, we will test this hypothesis in the context of these three areas (sample description, limitation, and data-supported claims). 
\begin{itemize}
\item H1: Recent articles in Phys. Rev.- PER tend to include sample description, include limitations on the study, and make data-supported claims.
\begin{itemize} 
\item H1a: Articles written during the beginning of PER did not tend to include sample description, limitations on the study, and make data-supported claims.
\end{itemize}
\end{itemize}

We will examine articles for the most recent 5 years of Phys. Rev.- PER (2012-2017). For older articles, we decided to use the creation of the Force Concept Inventory (FCI) as our starting point. The FCI was created in 1992. We will use a five- year span (1992-1997).  Because Phys. Rev.- PER did not exist then, we will  American Journal of Physics (AJP) and The Physics Teacher (TPT). Early volumes of AJP and TPT had some research papers including the FCI paper series.

\subsection{Analysis}
We will create a list of PR- PER articles that use the assessments we have listed. Then we will read the articles to see if the authors:
\begin{enumerate}
\item Describe the sample in terms of demographic information and institution context;
\item Discuss the limitations of their results in terms of population and are mindful that they do not generalize their results to all students/contexts; and
\item Make sure any claims are supported by the data.
\end{enumerate}

Each article will be examined for each potential issue. Each issue will be coded as the information is included or missing.

We will compare the most recent articles to ones in the past to determine whether H1 and H1a are supported.

\end{document}